# Multi-Level Steganography: Improving Hidden Communication in Networks

Wojciech Frączek, Wojciech Mazurczyk, Krzysztof Szczypiorski

Warsaw University of Technology, Faculty of Electronics and Information
Technology, Institute of Telecommunications, 15/19 Nowowiejska Str.
00-665 Warsaw, Poland
`wfraczek@gmail.com,{W.Mazurczyk, K.Szczypiorski}@tele.pw.edu.pl`

**Abstract.** The paper presents Multi-Level Steganography (MLS), which defines a new concept for hidden communication in telecommunication networks. In MLS, at least two steganographic methods are utilised simultaneously, in such a way that one method (called the upper-level) serves as a carrier for the second one (called the lower-level). Such a relationship between two (or more) information hiding solutions has several potential benefits. The most important is that the lower-level method steganographic bandwidth can be utilised to make the steganogram unreadable even after the detection of the upper-level method: e.g., it can carry a cryptographic key that deciphers the steganogram carried by the upper-level one. It can also be used to provide the steganogram with integrity. Another important benefit is that the lower-layer method may be used as a signalling channel in which to exchange information that affects the way that the upper-level method functions, thus possibly making the steganographic communication harder to detect. The prototype of MLS for IP networks was also developed, and the experimental results are included in this paper.

## 1  Introduction

Steganography is an art and science known for ages, whose main aim is to hide secret data (steganograms) in innocent-looking carriers [16]. The most suitable carrier is the one that is most commonly used (the use of such a carrier is not in and of itself an anomaly). Moreover, the modification of the carrier caused by inserting steganograms cannot be "visible" to the third party observer, i.e., he/she cannot point to the difference between modified and unmodified carrier if he/she is not aware of the steganographic procedure.

In telecommunication networks, all of the information hiding techniques that can be used to exchange secret data (steganograms) are called network steganography. Hidden communication network steganography utilises network protocols and/or relationships between them as a steganogram carrier [10]. It is important to emphasise that for a third party observer who is not aware of the steganographic procedure, the exchange of steganograms remains hidden. This is possible because inserting hidden data into a chosen carrier remains unnoticeable for users not involved in steganographic communication. Thus, not only the steganograms are hidden inside the carriers (network protocols), but because of the features of the carriers, the fact of the secret data exchange is also hidden. For review of the network steganography methods please refer to a survey by Zander et al. [19].

We wish to emphasise that network steganography can be utilised by decent users to exchange covert data, e.g., to circumvent censorship [2], to provide a communication channel between journalists and their information sources or by companies that are afraid of corporate espionage, but can also be used by intruders to leak confidential data or to perform network attacks [5, 18] . This is a usual trade-off that requires consideration in a broader steganography context, which is beyond the scope of this paper.

Each network steganography method may be characterised by three features: first, *steganographic bandwidth*, which describes how much secret data we are able to send using a particular method per time unit. Second, *undetectability* is defined as an inability to detect a steganogram inside certain carriers. The most popular way to detect a steganogram is to analyse statistical properties of the captured data and compare them to the typical properties of that carrier. The last feature is the *steganographic cost*, which describes the degree of degradation of the carrier caused by the steganogram insertion procedure. The steganographic cost depends on the type of the carrier, and if it becomes excessive, it leads to easy detection of the steganographic method. For example, if the method uses voice packets as a carrier for steganographic purposes in IP telephony, then the cost is expressed in conversation degradation. If the carri-er is certain fields of the protocol header, then the cost is expressed as a potential loss in that protocol functionality, etc.

For each method of network steganography, there is always a trade-off necessary between maximising steganographic bandwidth and still remaining undetected. A user can use a method naively and send as much secret data as is possible, but it simultaneously raises a risk of disclosure. Therefore, he/she must purposely resign from some fraction of the steganographic bandwidth in order to achieve undetectability.

Network steganography achieves security through obscurity; as long as the steganographic procedure remains unknown to third parties, it can be freely used to exchange hidden data. The problem arises when the functioning of the steganographic method is no longer secret. In such cases, anyone who is able to capture suspected traffic can extract and read hidden information (steganogram). One solution to this problem is to cipher steganograms, so in case of disclosure, it will not be readable. However, there is a question: how to exchange the cryptographic key? In overt communication, specialised key exchange protocols like Diffie-Hellman [17] can be utilised, but this is not an option for covert transmission, because such direct connection can look suspicious. Of course, one can always send it through covert channels, where the steganogram will be exchanged, but this approach raises two serious issues:

- The cryptographic key and ciphered steganogram is sent using the same steganographic method. Thus, the detection of this method results in the discovery of the cryptographic key and the steganogram content.
- Steganographic bandwidth devoted to carrying a user steganogram will be even more limited.

Another issue that steganographic communication must deal with is how to provide verification of the steganogram integrity after it is sent to the steganographic receiver. Usually, it will require sacrificing a fraction of the method's steganographic bandwidth to transmit additional data as well as a specialised protocol to be able to distinguish what is sent and when.

To address the abovementioned problems, in this paper, we propose to utilise a concept of Multi-Level Steganography (MLS), which was originally proposed by Al-Najjar for picture steganography in [1]. The idea in Al-Najjar's paper was to embed a decoy image into LSBs (Least Significant Bits) of the cover one and the real secret message is hidden into the LSBs of the decoy picture. Thus, the main application of MLS for digital image steganography was limited to make an extraction of the steganogram harder to perform.

We extend this concept for network steganography and redefine it to make it more general. MLS in telecommunication network is based on combining two or more steganographic methods in such a way that one method (the upper-level) is a carrier for the other method (the lower-level). The nature of network steganography environment i.e. binding of the overt communication process with steganographic method allows to pinpoint some useful MLS applications that can really improve hidden communications in telecommunication networks that were not considered before. The initial paper on MLS applied to IP networks was published by authors in [4]. This work significantly extends mentioned paper and its contributions are as follows:

- We provide detailed analysis of the potential MLS applications that can really improve hidden communication in telecommunication networks e.g. by providing means for cryptographic key exchange and/or steganogram integrity verification – both these issues as mentioned above are still an open challenges for network steganography.
- We develop a proof-of-concept, prototype implementation of two-method MLS for VoIP environment to prove that it is feasible.
- We present experimental results based on the MLS implementation to prove that it is useful for certain MLS applications provided above.

The rest of the paper is structured as follows. Section 2 introduces the concept of Multi-Level Steganography and its most important features. Section 3 describes, from our point of view, the most important applications of MLS. Section 4 presents first the implementation of MLS and experimental results. Finally, Section 5 concludes our work.

## 2 Multi-level Steganography (MLS) description and features

Multi-Level Steganography is a new concept of information hiding in telecommunication networks that uses features of an existing steganographic method (the upper-level method) to create a new one (the lower-level method). The idea of a simple two-method MLS, i.e., in which two steganographic methods are utilised as described above, and its comparison to the typical single network steganography method is presented in Fig. 1.

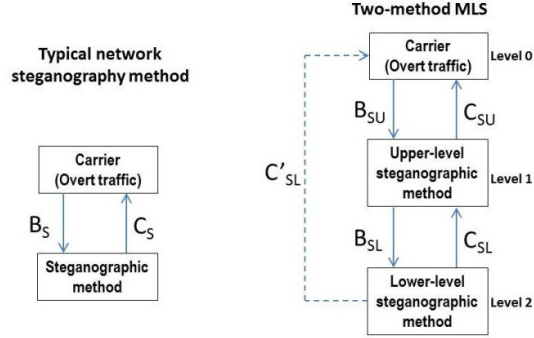

Fig. 1 The typical network steganography method (left) and the two-method MLS (right) comparison

In typical single-method network steganography, overt communication traffic is used as a carrier for secret data. By influencing the carrier, a certain steganographic bandwidth ($B_S$), which is defined as the amount of the steganogram transmitted using a particular method in one second ([b/s]), is achieved. However, the utilisation of $B_S$ may result in a certain steganographic cost ($C_S$) that expresses an impact (degradation) of a hidden data carrier due to steganographic procedure operations (see Section 1). The higher $B_S$ for given steganographic method we want to utilise the higher $C_S$ (the steganographic method has a greater impact on a hidden data carrier). If $C_S$ is excessive, then the detection of the method can be straightforward. Thus, a trade-off between $B_S$ and $C_S$ is always necessary.

As mentioned in Section 1, MLS is based on at least two steganographic methods. First, the upper-level method uses overt traffic as a secret data carrier. The second, the lower-level method, uses the way the upper-level method operates as a carrier. The indirect carriers for lower-level methods are still packets from overt communication, but the direct carrier is another (upper-level) method.

For the MLS case presented in Fig. 1, the upper-level method affects the carrier by introducing a certain cost $C_{SU}$, and under this circumstance, it achieves $B_{SU}$. The lower-level method relies on the upper-level one for its steganographic bandwidth $B_{SL}$. For this reason, the lower-level method can influence the upper-level one by introducing a cost $C_{SL}$ but also the overt communication by introducing a cost $C'_{SL}$. The cost $C'_{SL}$ depends on the choice of the lower-level method and, in particular, lower-level method can have no influence on the carrier i.e. it introduces no cost ($C'_{SL} \approx 0$).

## 2.1 General MLS scenario

In a more general scenario, MLS may be based on more than two steganographic methods; thus, more than two levels may be created (see Fig. 2, right). In Fig. 2, the MLS consists of 3 levels, so we use the terms Level 1 (or 2 or 3) steganographic method rather than upper- or lower-level to refer to each of them. Level 0 is considered as the overt channel. Of course, on each level, more than one steganographic method may be utilised; however, it may quickly degrade the carrier quality and thus make easy detection possible. The construction of an MLS has certain benefits compared to the scenario in which two (or more) unrelated steganographic methods are simultaneously utilised on the same carrier (Fig. 2, left):

- In general, the total steganographic cost of the MLS can be lower (for a given number of levels) than for the same number of methods used simultaneously on the same carrier (especially for the case where $C_{Sk0} \approx 0$ where $k>1$ is a number of levels in MLS).
- The detection of MLS is harder to perform because only the discovery of the higher level method can lead to the detection of the lower level methods.
- There is a direct relationship between the steganographic methods used for MLS construction. If some additional data are carried in lower level methods, this is a direct indication that it can be used for the benefit of the higher level method.

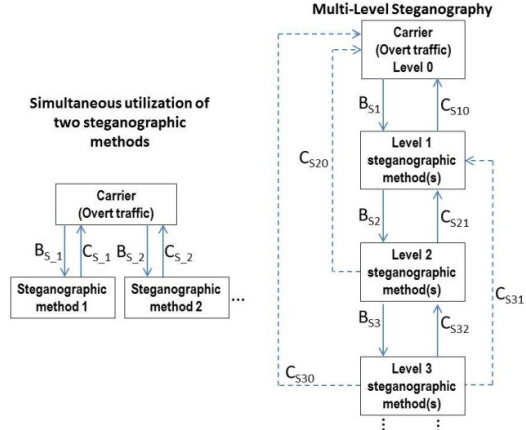

Fig. 2 General MLS case and simultaneous utilisation of multiple steganographic methods comparison

Thus, the entire steganographic bandwidth for MLS ($B_{S\text{-}MLS}$) from Fig. 2 (right) can be expressed as

$$B_{S-MLS} = \sum_{n=1}^{k} \sum_{m=1}^{l_n} B_{Snm}, \qquad (1)$$

where $k$ is the number of levels, $l_n$ is the number of steganographic methods used at level $n$ and $B_{Snm}$ is steganographic bandwidth of method $m$ at level $n$. If we assume that only one method can be used at each level, then Formula (1) can be simplified to

$$B_{S-MLS} = \sum_{n=1}^{k} B_{Sn}, \qquad (2)$$

where $B_{Sn}$ is the steganographic bandwidth of the method at level $n$.

The total cost is

$$C_{S-MLS} = \sum_{n=1}^{k} \sum_{o=1}^{l_n} \sum_{m=0}^{n-1} \sum_{p=1}^{l_m} C_{Snomp} < T, \qquad (3)$$

where $k$ is the number of levels, $l_n$ is the number of steganographic methods used at level $n$ and $C_{Snomp}$ is impact (cost) of method $o$ at level $n$ on the method $p$ at level $m$. Level 0 is level of overt communication and $l_0=1$ because there is only one overt channel. After reaching the threshold $T$, the steganographic method is easy to detect; thus, it is advised that $C_{S\text{-}MLS}$ should be always below it. In an ideal situation, the value of $C_{Snomp}$ equal to $0$ for $n>1$, which means that methods at level below $1$ have no impact on the upper-level ones or on the overt communication, and the total cost depends only on the level 1 methods. As in the formula for steganographic bandwidth, if we assume that only one method can be used at each level, then Formula (3) can be simplified to

$$C_{S-MLS} = \sum_{n=1}^{k} \sum_{m=0}^{n-1} C_{Snm} < T, \qquad (4)$$

where $C_{Snm}$ is the impact (cost) of the method at level $n$ on method at level $m$.

It is our view that two-method MLS is the most realistic scenario for network steganography and finding MLS scheme with more than two levels with satisfying steganographic bandwidth can be difficult to achieve. That is why in the rest of the paper, we limit our considerations to a two-method based MLS unless otherwise stated (and thus the terms upper- and lower-level methods when we refer to only two of them).

### 2.2 MLS features and hidden communication scenarios

MLS, in general, has two important features. First, the bandwidth of the lower-level method is a fraction of the bandwidth of the upper-level method. This is similar to the relationship between overt commu-nication bandwidth and upper-level steganography bandwidth (Fig. 3). The more redundancy and complexity in overt communication, the more hidden data can be inserted and exchanged covertly.

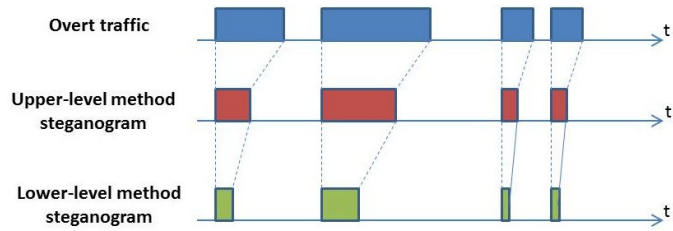

Fig. 3 Upper- and lower-level steganographic bandwidth

Second, the lower-level method is potentially harder to detect than the upper-level one. It results from the fact that the lower-level method functioning entirely depends on upper-level one. Thus, the adversary has to detect the upper-level method first in order to look for the lower-level one. Moreover, the undetectability of MLS may be the same, greater or lower than as if only an upper-level method were used, depending on the choice of the upper- and lower-level methods.

The introduction of MLS influences also possible steganographic communication scenarios (Fig. 4), which were introduced in [19]. In this paper, there are four communication scenarios, which depend on whether the sender and receiver of the steganograms are the sender and receiver of overt communication or middlemen.

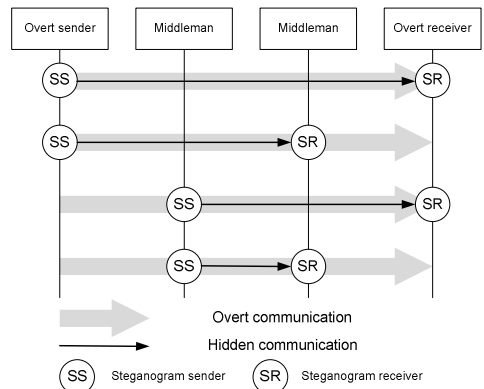

Fig. 4 Communication scenarios in typical, single-method network steganography

In Multi-Level Steganography, the number of communication scenarios increases to 16. Each scenario from Fig. 4 can be replaced with four new scenarios, which are presented in Fig. 5.

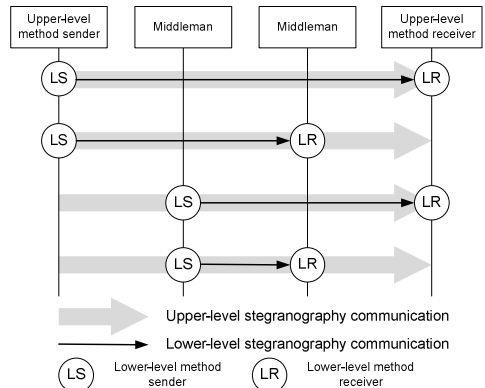

Fig. 5 MLS communication scenarios

The upper-level method steganogram sender and receiver in Fig. 5 are the steganography sender and steganography receiver, in Fig. 4, respectively. The middlemen in Fig. 4 are network devices that are placed between the sender and receiver of the upper-level steganographic method. They have to be aware of the upper-level method in order to use it to create the lower-level method.

Each known steganographic method can presumably be used as an upper-level method. The main problem is to find a suitable lower-level method that will cooperate with the upper-level one. If we consider, for example, methods like LACK [13], RSTEG [12] or SCTP hybrid method [3] as an upper-level method (which modify the packet payload to insert a steganogram), then the lower-level method can hide bits in the number of packets that the steganogram of the upper-level method sends in established time periods. For example, if a packet with the steganogram is sent during a one-second period – it means binary "1", else (i.e. if packet is sent later) – it means binary "0".

## 3. MLS Applications

Multi-Level Steganography can be utilised to achieve various aims – it all depends on how it will be used. Here we present several of the most interesting MLS applications, in our opinion. The benefits of MLS for hidden data exchange are summarised in Table I.

Table I – MLS benefits and possible applications

| MLS benefit | Described MLS application |
| --- | --- |
| Increased steganographic bandwidth for user data | Using two or more steganographic methods increases the total steganographic bandwidth achieved for user data compared with a single steganographic method. |
| Increased undetectability | An upper-level method controlled by information carried by the lower-level method (Sec. 3.1). |
| Steganogram transmission reliability | Lower-level method carrying information for steganogram integrity verification (Sec. 3.1). |
| Harder steganogram extraction and analysis | 1. Cryptographic key carried by lower-level method and upper-level method steganogram ciphered (Sec. 3.1).<br>2. Parts of the steganogram sent using the upper-level and others by the lower-level method (Sec. 3.3).<br>3. Steganogram carried only by the lower-level method; upper-level steganogram only for masking (Sec. 3.2). |
| Steganographic cost unchanged | In best-case scenario, depends on the upper- and lower-level methods used, but can be the same as for utilisation of the upper-level method alone. |

Let us consider the abovementioned MLS applications based on where the steganogram is inserted. There are three possible cases:
- Steganogram is carried only by upper-level method
- Steganogram is carried only by lower-level method
- Steganogram is carried by both upper- and lower-level methods

The rest of this section describes these cases in detail.

**3.1 Only upper-level method carries steganogram**

If steganogram is only carried by the upper-level steganographic method, then the lower-level method may be utilised for special purposes. For example, it may carry a cryptographic key that was used to cipher the steganogram sent by the upper-level method (Fig. 6). Such an MLS application is possible only for steganographic communication scenarios in which the upper- and lower-level hidden communication paths are the same (compare Fig. 4 and 5). It is assumed that the steganogram was ciphered before initiating the steganographic exchange to avoid any unnecessary delays. When overt traffic is sent, so are the parts of the ciphered steganogram (upper-level method) and bits of the cryptographic key (lower-level method). After all bits of the ciphered steganogram and cryptographic key are successfully received, the steganographic receiver is ready to decipher the steganogram. If the hidden communication, for some reason, did not last long enough to send the whole cryptographic key, then the steganographic receiver stores ciphered steganogram and waits for the next covert communication from this steganographic sender.

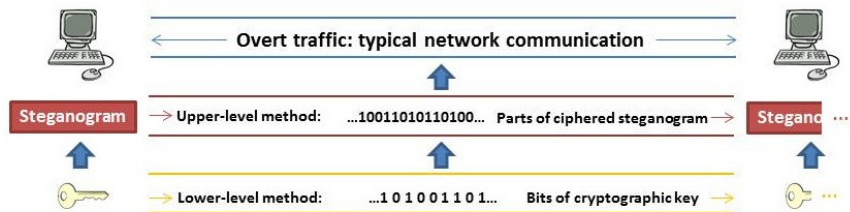

Fig. 6 MLS application: lower-level method carries cryptographic key

Another application of Multi-Level Steganography may be to use lower-level method steganographic bandwidth to verify the integrity of the steganogram carried by the upper-level method (Fig. 7). Before the steganographic exchange begins, a hash function ($H$) is used to calculate a hash, which is then transferred to the steganographic receiver by the lower-level method. After the hidden data transmission ends, the receiver calculates the corresponding hash on the received and extracted steganogram. Next, the locally calculated hash is compared to the received one. If they are the same, then the transmission was successful. If not, then some transmission error has occurred and the steganogram must be resent sometime in the future.

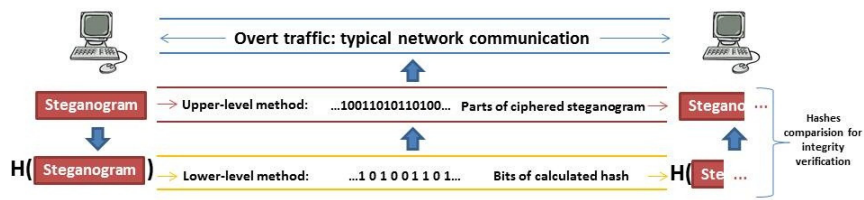

Fig. 7 MLS application: lower-level method carries steganogram integrity information

The integrity of the steganogram may be also verified in a different way if the upper- and lower-level methods are chosen properly. If we divide the steganogram carried by the upper-level method into parts and assign them sequence numbers, then the lower-level method may be used to transfer the sequence number of the corresponding part of the steganogram sent the by upper-level one.

Another interesting application of Multi-Level Steganography is to use information carried by lower-level method to control the way the upper-level method works. Such functionality can be helpful to limit the chance of disclosure, e.g., by changing characteristic features of the particular steganographic method during hidden data exchange (Fig. 8). Some methods allow their behaviour to be changed while transmitting steganograms. One example of such method is the Cloak method [11], which uses many TCP flows between a steganographic sender and receiver to enable secret communication. In Cloak, one can change its parameters during the steganographic data exchange. There are two parameters that can be changed: the numbers of segments ($N$) and TCP flows ($X$). The problem with modifying them during covert transmission is how to indicate to the steganographic receiver that these parameters have changed and the steganogram is inserted elsewhere compared to the beginning of transmission. The solution to this problem may be MLS. When overt communication begins, the upper-level method starts to send steganograms in a predetermined mode. At the same time, the lower-level method (that can be analogous to the lower-level method provided in prototype MLS implementation in Section 4) is utilised to transfer control information that affects the mode of the upper-level method (Fig. 8, case 1). After the lower-level method succeeds in its transmission, the steganographic receiver acknowledges reception of the new parameters, and from now on, the sender incorporates the changes into the steganographic procedure (Fig. 8, case 2). This effect, continuously repeated while the hidden data exchange lasts, makes detection more difficult.

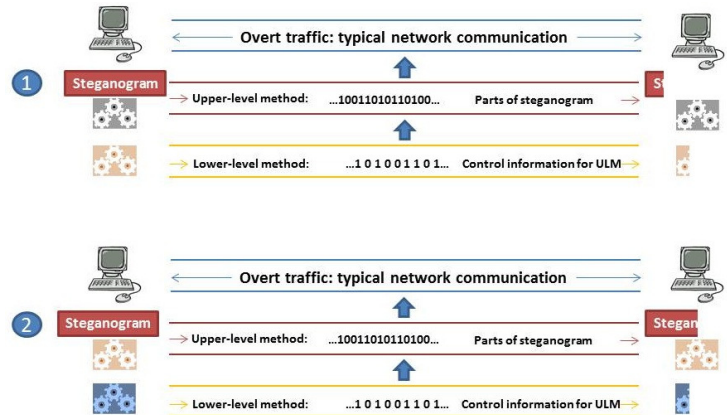
Fig. 8 MLS application: lower-level method as a signalling channel to control upper-level method

Moreover, if the steganographic bandwidth of the lower-level steganographic method is sufficient, all of the abovementioned applications of MLS may be used at once, i.e., the lower-level method can carry cryptographic key and control information for the upper-level one. However, this requires the existence of some protocol for lower-level method that will govern what and when is sent to the steganographic receiver.

### 3.2 Only lower-level method carries steganogram
In this case, the upper-level method also carries some information, but it is not used to transfer steganograms. Its purpose is to mask the existence of the lower-level method, but it does not have any meaning. The real steganogram is carried only by the lower-level steganographic method. If the upper-level steganographic exchange is uncovered, then the secret information will not be compromised. However, the main disadvantage of this solution is its limited steganographic bandwidth, which for large volumes of data may limit its usage.

### 3.3 Both upper- and lower-level methods carry steganogram
More interesting than the previous case is the situation in which the steganogram is carried using the upper- and lower-level methods. There are two possibilities, based on whether or not these steganograms are related:
- The upper- and lower-level methods send separate steganograms that are not related,
- The original steganogram is divided into pieces, and some pieces are sent using the upper-level and some using the lower-level method.

In the first case, the lower-level method serves as a separate steganographic channel in which additional secret data can be exchanged. In the second case, the original steganogram is divided into two parts; the first part is sent using the upper-level method and the second part using the lower-level one (Fig. 9, case 1). The original steganogram is successfully received and can be read if it is extracted from both level methods and properly combined (Fig. 9, case 2). Such a simple operation can make the steganogram harder for a third party observer to extract and analyse. If he/she is aware of the existence of only the upper-layer method, then he/she can extract only parts of the shattered steganogram.

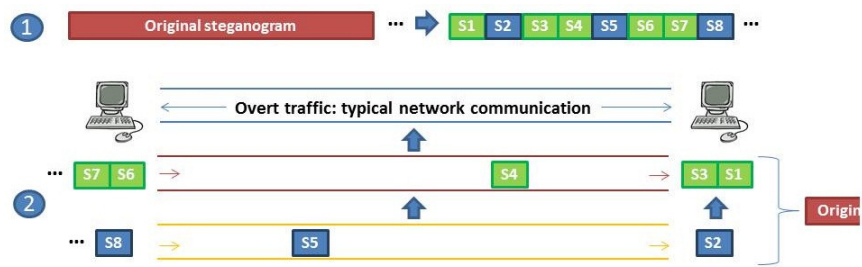
Fig. 9 MLS used for divided steganogram transmission using inter-leaving

# 4. MLS Prototype and Experimental Results

For MLS prototype development two steganographic methods were used. As an upper-layer method, LACK [13] was utilised. This method is intended for a broad class of multimedia, real-time applications like IP telephony. In IP telephony, conversation is based on exchanging RTP (Real-Time Transport Protocol) streams between calling parties. Each of the RTP is uniquely identified with a sequence number.

LACK utilises the fact that for usual multimedia communication protocols like RTP, excessively delayed packets are not used for the reconstruction of transmitted data at the receiver, i.e., the packets are considered useless and discarded. The idea of LACK is as follows. At the transmitter, some selected audio packets are intentionally delayed before transmitting. If the delay of such packets at the receiver is considered excessive, the packets are discarded by a receiver that is not aware of the steganographic procedure. The payload of the intentionally delayed packets is used to transmit secret information to receivers aware of the procedure, so no extra packets are generated. For unaware receivers, the hidden data are "invisible".

A lower-level method is based on proper RTP sequence number matching. It modifies the choice of the RTP packet (its sequence number) used for LACK purposes depending on the steganogram bits to be sent.

The functioning of the implemented MLS prototype is presented in Fig. 10. First, due to the LACK method, a RTP packet is selected for steganographic purposes (1). If the RTP sequence number is not suitable for the lower-level method, then one of the neighbouring RTP packet is selected instead with a suitable sequence number (2). Next, the chosen packet is delayed at the transmitter and then sent through the communication channel to the receiver, and the original payload is replaced with the steganogram (3). At the receiver, the LACK packet was considered lost; thus, when it comes, it is not used for voice reconstruction. Instead, the payload of the RTP packet is extracted and treated as an upper-level steganogram, and based on this packet sequence number, a lower-level steganogram is also determined (4).

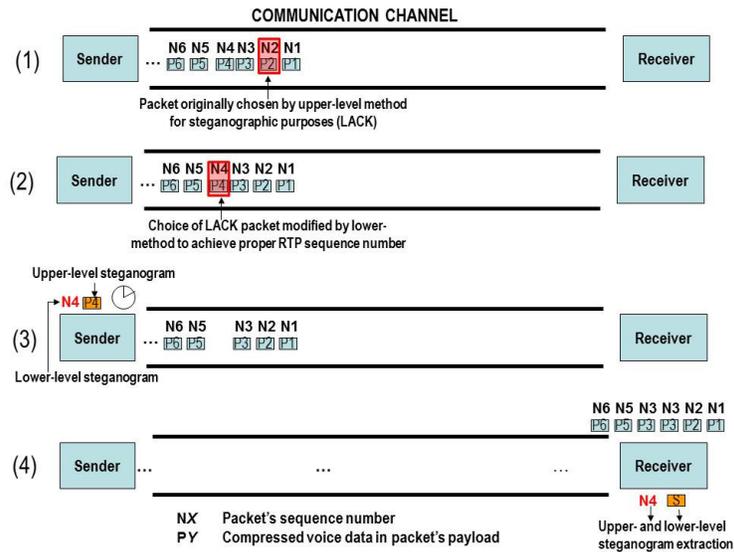

Fig. 10 MLS prototype functioning

In the simplest case, we can assume that an odd RTP sequence number of LACK packet means a binary "1", and an even RTP sequence number means a binary "0". For example, if a user has to send bits "101", then the sequence number of first LACK packet will be odd, the sequence number of the second LACK packet will be even and the sequence number of third LACK packet will be odd. This method can be further extended to convey more than 1 bit per LACK packet. Steganogram bits can be determined by the last x bits of sequence number of the LACK packet. Then, the bandwidth of this method is x bits/LACK packet, since each LACK packet carries x bits in its sequence number. However, it is important not to change the order of RTP packets when the LACK packet is changed due to the influence of the lower-level method, because it may lead to errors in the steganography data received.

### 4.1. MLS Prototype Implementation

Implementation of MLS prototype was based on *MjSip* [14] project. It is a Java implementation of VoIP softphone based on a SIP (Session Initiation Protocol) signalling protocol. Only the user agent application was utilised; the SIP server was omitted because it does not affect the results of experiments (the RTP streams are exchanged directly between end users, without using the SIP server). In the SIP User Agent application, a simple PLC (Packet Loss Concealment) method was implemented, as softphones usually have some way to deal with packets losses. PLC mechanisms are used to limit quality degradation due to packet loss – in the simplest scenario, these insert a repetition of the last received packet to substitute for a missing one [8]. This PLC method was added to the SIP User Agent application.

The implementation of the upper-level method, LACK, was straightforward. For each RTP packet chosen for LACK purposes, the payload consists of two parts: steganogram and hash. The hash is computed for the steganogram carried in that packet using the MD5 (Message Digest 5) hash function. It allows the receiver to distinguish LACK packets from normally transmitted non-steganographic ones.

Two parameters of the LACK method were affected: the probability that a packet is used for LACK purposes ($p_{LACK}$) and the minimum delay of LACK packets. For each RTP packet, a pseudo-random number between 0 and 1 was generated, and it was tested whether the number is smaller than established probability of sending a LACK packet. If this was the case, then the packet was chosen for steganographic purposes.

The implementation of the lower-level method required (if necessary) modification of the upper-level choice of LACK packet. If the upper-level method selected an RTP packet for LACK purposes whose sequence number satisfied the needs of the lower-level method, i.e., the steganogram bits to be sent, the packet is not changed. Otherwise, an RTP packet with the proper RTP sequence number is selected. The lower-level method tries to select RTP packets as close as possible to the packets originally selected by the upper-level method.

For example, let us assume that the lower-level method to carry steganogram utilizes two least significant bits of each RTP packet's (selected by upper-level method) sequence number. In that case, if the bits of the lower-level steganogram to be sent are "10" and the sequence number chosen for LACK purposes is odd e.g. it is 51 (110011), then lower-level method influence the choice by changing it to the neighbouring RTP packet with even sequence number e.g. 54 (110110). Then this RTP packet's payload will be replaced with the upper-level steganogram.

The main issue was not to change the order of RTP packets because of a change in the LACK packets imposed by the lower-level method. The problem was solved by marking only one packet for LACK purposes at any given moment. If more than one packet was chosen to be marked for LACK purposes at the same time, then only one was marked. However, it was noted that additional packets must be sent as LACK packets. For example, if the RTP sequence number of the packet chosen by upper-level method is 51 but was changed to 54 by the lower-level method, and, simultaneously, sequence number 53 was the next chosen by upper-level method, then the packet with sequence number 53 is not considered as a LACK packet (in order not to break the rule "one packet for LACK purposes at any given moment"). After sending the LACK packet (with original sequence number 51), additional packet will be chosen for LACK purposes.

One parameter of the lower-level method was subject to configuration, which was the number of lower-level bits of steganogram that are sent with each LACK packet. This parameter must be set the same for both the sender and receiver of the lower-level steganogram.

### 4.2. Experiment Methodology and Results

The experimental setup is presented in Fig. 11. The environment for experiment was a LAN network, so no packets were lost or excessively delayed except intentionally, which permitted us to evaluate the sole impact of LACK and MLS on voice quality, without any network-related or endpoint-related interferences.

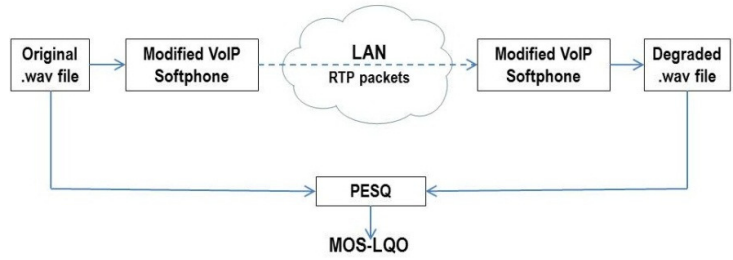
Fig. 11 MLS experimental setup

The conversation was recorded in advance and encoded with G.711 (each RTP packet carries 20 ms of voice using 160 bytes; the packet flow rate is 50 packets per second) and then saved as an input .wav file. The duration of the conversation was set to 9 minutes, as it is experimentally verified that the average call duration for IP telephony is in the range 7-11 minutes [6]. Then, parts of the .wav file were inserted into the payloads of consecutive RTP packets. Next, the RTP stream was influenced by the chosen steganographic methods:

- Only LACK, where MLS is not used – this case will be treated as a reference for MLS.
- MLS that transfers 1 bit of the lower-level method steganogram in a single LACK packet (MLS-1).
- MLS that transfers 2 bits of the lower-level method steganogram in single LACK packet (MLS-2).
- MLS that transfers 3 bits of the lower-level method steganogram in a single LACK packet (MLS-3).

In the next step, the RTP stream was sent to the receiver, which reconstructed the voice conversation and saved it to the output .wav file. Then, the parts of about 30 s length of original (input) and degraded (output) .wav files were compared using the PESQ method [9], and the MOS-LQO (Mean Opinion Score - Listening Quality Objective) value was obtained. Then the average MOS-LQO was calculated. By performing experiments in a strictly controlled environment with no losses and excessive delays, we were able to assess the real influence of MLS on the conversation quality. For each steganographic method mentioned above, the experiment was repeated 10 times, and the average results are presented.

We decided to set the probability of selecting an RTP packet for steganographic (LACK) purposes by upper-level method to 0.032 because it resulted in MOS-LQO value around 3.6 which is regarded as a quality compared with that achieved in PSTN networks. It also means that the cost of the upper level method is $C_{SU} \approx 0.7$ in MOS scale (see Fig. 1), because the quality of the G.711-based connection without LACK is about 4.3.

The probability of selecting an RTP packet for steganographic purposes by upper-level method was set the same for all experiments, but it was not always achieved. The actual, real value of the RTP packets selected for LACK purposes could be different because of process of generation random numbers.

For presented experimental setup we measured steganographic bandwidth of upper- and lower-level methods ($B_{SU}$, $B_{SL}$) and the corresponding costs introduced ($C_{SU}$, $C_{SL}$). Obtained experimental results are presented in Table II and Fig. 12 and 13.

Table II Experimental results

|  | LACK | | MLS-1 | | MLS-2 | | MLS-3 | |
|---|---|---|---|---|---|---|---|---|
|  | Average | CI (95%) | Average | CI (95%) | Average | CI (95%) | Average | CI (95%) |
| MOS-LQO | 3.609 | 0.018 | 3.617 | 0.025 | 3.626 | 0.015 | 3.625 | 0.019 |
| $C_{SU}$ [MOS] | $\approx 0.7$ | | $\approx 0.7$ | | $\approx 0.7$ | | $\approx 0.7$ | |
| $C_{SL}$ [MOS] | N/A | | $\approx 0$ | | $\approx 0$ | | $\approx 0$ | |
| $p_{LACK}$ | 0.0317 | 0.0008 | 0.0323 | 0.0017 | 0.0316 | 0.0007 | 0.0315 | 0.0006 |
| $B_{SU}$ [bit/s] | 1827.41 | 48.39 | 1837.87 | 44.13 | 1822.08 | 39.20 | 1812.48 | 36.74 |
| $B_{SL}$ [bit/s] | 0.00 | 0.00 | 1.60 | 0.04 | 3.16 | 0.07 | 4.72 | 0.10 |

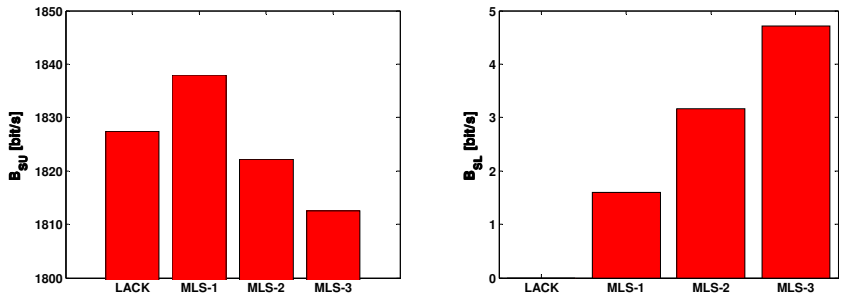

Fig. 12 Steganographic bandwidth of upper- (left) and lower-level (right) methods

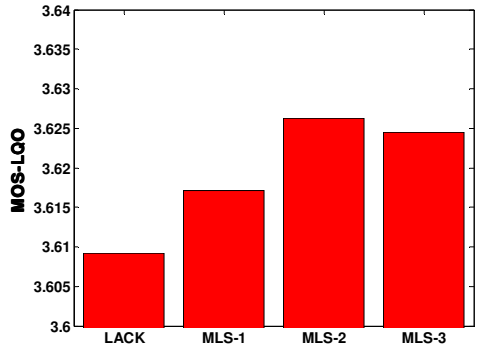

Fig. 13 Experimental voice quality results (MOS-LQO)

Experimental results prove that for the MLS prototype presented, which uses LACK as an upper-level method, adding a lower-level steganographic method has negligible impact on the voice quality as well as on upper-level steganographic bandwidth, thus the cost $C_{SL} \approx 0$. It is also worth noting that for the presented prototype implementation and chosen upper- and lower-level methods the $C_{SL}' \approx 0$ as there is also no lower-level method's direct influence on VoIP conversation.

When LACK is used alone during an IP telephony conversation, its steganographic bandwidth is about 1830 bit/s, after applying MLS, it remains in the range 1810-1840 bit/s (Fig. 10, left). The difference in the LACK steganographic bandwidth comes from the fact that it was hard to obtain $p_{LACK}$ precisely in the experiments for MLS-1, -2 and -3 (Table II). In these circumstances, for the lower-level MLS method, a steganographic bandwidth was obtained of up to 5 bit/s (MLS-3). The voice quality scores achieved were also similar and in the range of 3.61-3.63.

Of course, such high values of upper-level steganographic bandwidth are achieved when LACK introduces about 3% packet losses. In real-life IP networks, causing so many losses may have a great impact on voice quality because they are added to the network and jitter buffer losses. For example, for G.711 the maximum packet loss acceptable is 5% (when PLC was used) [15]. Thus, if LACK is to remain undetected, the losses it introduces should be kept at a reasonable level.

However, even if fewer losses are introduced, the resulting steganographic bandwidth of the lower-level method may be sufficient for some MLS applications outlined in Table I. Earlier experimental research for IP telephony verified that the average call duration falls in the range of 7-11 minutes [7]. This means that for typical 9 minute call from 540 (MLS-1) to 2700 (MLS-3) bits can be transferred using lower-level steganographic method. Thus, the lower-level method's steganographic bandwidth is suitable in the proposed MLS prototype to provide reliability of the upper-level steganogram (by carrying integrity hash) or to carry a cryptographic key if one is needed, thus making it harder to extract and analyse an upper-level steganogram.

Obviously, the increase in total steganographic bandwidth is not significant, and is about 0.3%, thus making it hard to use applications of MLS where parts of the steganogram are sent using the upper-level and others by the lower-level method or the steganogram is carried only by the lower-level method and the upper-level steganogram only for masking.

However, it must be noted that for the MLS prototype developed, the steganographic cost is unchanged when compared to the situation when only LACK is used. To conclude, what MLS applications, out of those mentioned in Table I, may be applied in a particular case depends on the choice of the upper- and lower-level methods – different MLSs may be more suitable for different MLS applications.

## 5. Conclusions and future work

In this paper, a new concept for performing hidden communication, called Multi-Level Steganography for network steganography, was presented. MLS consists of at least two steganographic methods are utilised simultaneously, in such a way that one method (called the upper-level) serves as a carrier for the second one (called the lower-level). Such relationship between two (or more) information hiding solutions has several potential benefits, e.g., it may provide increased steganographic bandwidth or increase undetectability. It can also be utilised for ensuring the reliability of steganogram transmission reliability or making steganogram extraction and analysis harder to perform.

The nature of network steganography environment i.e. binding of the overt communication process with steganographic method allows to pinpoint some useful MLS applications that can really improve hidden communications in telecommunication networks that were not considered before.

The MLS prototype was developed to prove its concept. It was implemented in an IP telephony environment. It was based on a previously introduced LACK solution as an upper-level method and selecting odd/even RTP sequence numbers for LACK packets as a lower-level method. For this prototype, the steganographic cost was unchanged compared to the situation when only the upper-level method was used.

Experimental results were obtained that proved that some of the described above MLS applications can be easily applied. It turned out that the lower-level method's steganographic bandwidth is suitable to provide reliability of the upper-level steganogram (by carrying integrity hash) or to carry a cryptographic key that secret data was encrypted with, thus making it harder to extract and analyse an upper-level steganogram. It can be also utilised to increase the upper-level method undetectability by utilizing the lower-level method's steganographic bandwidth to exchange control information between the covert parties that will influence the upper-level one functioning.

In general, for MLS, the most important is the choice of upper- and lower-level methods. In an ideal situation, the steganographic cost of the lower-level method should be equal to 0 or very small when compared with the steganographic cost of the upper-level one. Moreover, the higher the steganographic bandwidth of the lower-level method, the more the described MLS applications can be applied.

Our future work will be focused on developing more efficient MLS schemes because the benefits for such constructions of hidden data exchange are considerable and they solve some open challenges related to network steganography (providing steganogram reliability and cryptographic key exchange). Future work will also include analysing MLS detection options.

**ACKNOWLEDGMENT**
This work was supported by the Polish Ministry of Science and Higher Education and Polish National Science Centre under grants: 0349/IP2/2011/71 and 2011/01/D/ST7/05054.